\documentclass[
twocolumn,
prl,showpacs,preprintnumbers,amsmath,amssymb]{revtex4}

\usepackage{graphicx}
\usepackage{amssymb}
\usepackage{dsfont}
\usepackage{bm}

\begin{document}

\title{Intrinsic instability of sonic white holes}

\author{U. Leonhardt$^{1}$}
\author{T. Kiss$^{1,2,3}$}
\author{P. \"Ohberg$^{1,4}$}

\affiliation{$^{1}$School of Physics and Astronomy, University of St
  Andrews, North Haugh, St Andrews, KY16 9SS, Scotland }
\affiliation{$^{2}$Research Institute for Solid State Physics and Optics,\\ 
H-1525 Budapest, P.~O.~Box 49,  Hungary}
\affiliation{$^{3}$
Institute of Physics, University of P\'ecs,
Ifj\'us\'ag u.\ 6. H-7624 P\'ecs, Hungary}
\affiliation{$^{4}$Department of Physics, University of Strathclyde,\\
Glasgow G4 0NG, Scotland}             

\begin{abstract}
Artificial black holes, such as sonic holes in Bose-Einstein condensates,
may give insights into the role of the physics at the event horizon 
beyond the Planck scale.
We show that sonic white holes give rise to a discrete spectrum 
of instabilities that is insensitive to the analogue of 
trans-Planckian physics for Bose-Einstein condensates.
\end{abstract}

\pacs{03.75.Fi, 04.70.Dy}

\maketitle

Hawking predicted \cite{Hawking} that black holes generate thermal
radiation due to quantum effects \cite{BirrellDavies,Brout}.
Hawking's effect bridges two vastly different areas of the physical
sciences --- cosmology and quantum mechanics, but unfortunately, the
effect is too feeble to be observable for the known solar-mass or
larger black holes.  However, condensed-matter or optical analogs may
be able to demonstrate the equivalent of Hawking radiation in the
laboratory
\cite{Book,Unruh,Visser,Corley,JV,CJ,Garay,Barcelo,Laval,Fischer,Chapline,Dielectric,Leo}.
In understanding the mechanism of laboratory analogs one could perhaps
gain insight into the anatomy of genuine quantum black holes.  For
example, Hawking's effect appears to rely on a theory that predicts
its own demise \cite{TransPlanckProblem} --- radiation outgoing from
the event horizon seems to originate from wavelengths beyond the
Planck scale where the physics is unknown.  Yet the theory of
condensed-matter analogues indicates that Hawking's effect is robust
against trans-Planckian physics \cite{TransPlanck}.  In this Letter we
show that white holes with effective surface gravity $\alpha$ give
rise to a discrete spectrum of dynamical instabilities with decay
constants
\begin{equation}
  \label{eq:result}
  \gamma = 2n\alpha \,,\quad n \in \mathbb{Z} \,,
\end{equation}
whereas black holes may be stable. According to Corley and Jacobson
\cite{CJ}, a black-hole white-hole pair should act as a black-hole
laser, {\it i.e.} as an amplifying medium for Hawking radiation with
the two horizons forming a resonator, giving rise to a dynamic
instability of Hawking radiation.  This instability has been
attributed to physics beyond the Planck scale.  Here we point out that
the discrete spectrum of a single white hole is insensitive to the
equivalent of trans-Planckian physics within the model we have
employed, the sonic hole in a Bose-Einstein condensate.  Such a model
may be regarded as the {\it drosophila} of the artificial black holes
\cite{Garay,Laval}, because it is the simplest system to study
theoretically and it is within experimental reach.

Sonic holes are inspiring, because of the mathematical equivalence
between the propagation of sound in fluids and of scalar waves in
general relativity \cite{Unruh,Visser}. Consider sound waves in an
irrotational fluid of density $\rho_0$ and flow $\mathbf{u}$. The
velocity potential $\varphi$ and the density perturbations $\rho_s$ of
sound obey the linearized equation of continuity and the linearized
Bernoulli equation \cite{LL6}
\begin{eqnarray} 
 \partial_t \rho_s + \nabla \cdot (\mathbf{u} \rho_s + \rho_0 \nabla
 \varphi ) &=& 0\,, \label{eq:cont} \\
 (\partial_t + \mathbf{u} \cdot \nabla ) \varphi + c^2
 \frac{\rho_s}{\rho_0} &=& 0\,. \label{eq:bern}
\end{eqnarray}
The resulting wave equation in relativistic notation reads
\begin{equation}
  \label{eq:wave}
  D_\nu D^\nu \varphi = \frac{1}{\sqrt{-g}} \, \partial_\mu \sqrt{-g} \, g^{\mu
  \nu} \partial_\nu \varphi = 0
\end{equation}
with the effective space-time metric \cite{Visser}, in $d$ spatial
dimensions,
\begin{equation}
  g_{\mu \nu} = \left( \frac{\rho_0}{c^3} \right)^{2-2 d} 
  \left(
    \begin{array}{cc}
      c^2 - u^2& \mathbf{u}\\
      \mathbf{u} & - \mathds{1}
    \end{array}
    \right) \, .
\end{equation}
Nonuniform profiles of $\rho_0$, $c^2$ and $\mathbf{u}$ may generate
effective space-time geometries that are sufficiently rich to possess
event horizons \cite{Book}. At a sonic horizon the flow exceeds the
speed of sound.
Sound waves propagating against the current freeze and, in turn, 
their wavelenghts shrink dramatically.
We are interested in effects dominated by this wave catastrophe \cite{Leo}.
In this case we can use the simple one-dimensional model
\begin{equation}
  \label{eq:model}
  u = -c + \alpha z \, .
\end{equation}
Here $z$ denotes the spatial coordinate orthogonal to the horizon at
$z=0$, $\alpha$ characterizes the surface gravity or, in our
acoustic analog, the gradient of the transonic flow, and
$\rho_0$ and $c$ are assumed to be constant. 
Strictly speaking, we should complement the flow component 
(\ref{eq:model}) in the $z$ direction by appropriate components 
in the $x$ and $y$ directions,
in order to obey the continuity of the flow.
But as long as we focus on effects on length scales smaller than
$|c/\alpha|$ we can ignore the other dimensions of the fluid.
Equations (\ref{eq:cont}) and (\ref{eq:bern}) have the solutions
\begin{equation}
  \label{eq:sol}
  \varphi = \varphi_0 (\tau) \, , \quad \rho_s = \frac{\rho_0}{c \alpha z}
  \frac{d \varphi}{d \tau} \, , \quad \tau = \frac{ \ln
  (z/z_\infty)}{\alpha} - t \, ,
\end{equation}
with the arbitrary function $\varphi_0$ and the constant $z_\infty$,
describing wavepackets propagating against the current. Such
wavepackets are confined to either $z<0$ or $z>0$, depending on the
sign of $z_\infty$, which indicates that the place where the flow
exceeds the speed of sound, $z=0$, indeed establishes the acoustic
equivalent of the event horizon. Depending on the sign of $\alpha$,
two cases emerge: the sonic black hole and the white hole
\cite{Garay}.
The black hole is characterized by a positive
velocity gradient $\alpha$, the flow goes from subsonic to supersonic
velocity at the horizon, whereas $\alpha$ is negative for the white
hole \cite{Garay}
where the flow slows down from supersonic to subsonic speed. No
sound wave can leave  the supersonic zone of the black hole and no
sound can enter the white hole.

Experimental tests of the subtle quantum effects of the acoustic
horizon \cite{Book,LKOReview} will take the best superfluids currently
available, Bose-Einstein condensates of dilute gases
\cite{Anglin,Dalfovo}. Moreover, a condensate is one of the
simplest physical systems to draw theoretical analogies between the
quantum mechanics of fluids and quantum effects of gravity
\cite{Book}. The mean-field wavefunction $\psi_0$ of the condensate
\cite{Dalfovo} represents an irrotational fluid,
\begin{equation}
  \psi_0 = \sqrt{\rho_0}\, e^{i S_0} \, , \quad \mathbf{u} =
  \frac{\hbar}{m} \nabla S_0 \, ,
\end{equation}
where $m$ denotes the atomic mass. Sound waves are perturbations
$\psi-\psi_0$ of the fluid, with
\begin{equation}
  \label{eq:hydro}
  \psi = \sqrt{\rho}\, e^{i S} \, , \quad \rho = \rho_0 + \rho_s \, ,
  \quad S=S_0 + \frac{m}{\hbar} \varphi \, .
\end{equation}
Sound quanta, phonons, are elementary excitations characterized by the
Bogoliubov modes $u_n$ and $v_n$ \cite{Dalfovo,Fetter}, with 
\begin{equation}
  \label{eq:psiuv}
  \psi = \psi_0 + e^{i S_0} (u_n + v_n^*) \, .
\end{equation}
We determine the modes from the solution (\ref{eq:sol}) of the
hydrodynamic equations (\ref{eq:cont},\ref{eq:bern}) by comparing the
representations (\ref{eq:hydro}) and (\ref{eq:psiuv}) in the limit of
$|\rho_s/\rho_0|\ll 1$, $|\varphi/S_0|\ll \hbar/m$ \cite{LKO}. We find
the single-frequency Bogoliubov modes of the transonic flow
(\ref{eq:model}),
\begin{eqnarray}
  \label{eq:uv}
   u_n &=& A_n \left( \frac{\omega}{2 \alpha z} + \frac{m c}{\hbar}
  \right) z^{i\omega /\alpha} e^{-i \omega t} \, , \nonumber \\
   v_n &=& A_n \left( \frac{\omega}{2 \alpha z} - \frac{mc}{\hbar}
  \right) z^{i \omega/\alpha} e^{-i \omega t} \, .
\end{eqnarray}
Consider the analytic continuation of $z^{i \omega /\alpha}$ to
complex $z$. Suppose that $z^{i \omega/\alpha}$ is analytic on either
the upper $(+)$ or lower $(-)$ half plane. Consequently, we get for
real and positive $z$
\begin{equation}
  \label{eq:analytic}
  (-z)^{i \omega/\alpha} = e^{- \pi (\pm \omega /\alpha)} z^{i \omega
  / \alpha} \, .
\end{equation}
The Bogoliubov modes differ on the two sides of the horizon, depending
on the analyticity of $z^{i \omega /\alpha}$, which reflects the
independence of the two sides of the sonic horizon.

Transonic flows are notoriously plagued by hydrodynamic instabilities
\cite{LL6}, unless they are generated in appropriately designed
nozzles such as the Laval nozzle \cite{Courant} of a rocket engine.
Consider unstable elementary excitations corresponding to Bogoliubov
modes with complex frequencies $\omega$ \cite{LKO},
quasinormal modes \cite{Unstable}. The dynamic
equations of the excitations, the Bogoliubov-deGennes equations
\cite{Dalfovo,Fetter,LKO}, possess a four-fold symmetry in the complex
frequency plane \cite{Garay,LKO}: If $\omega$ is the frequency of a
solution then there exist solutions for the frequencies
$\omega^*$,$-\omega$ and $-\omega^*$ as well. The Hermiticity of the
underlying many-body Hamiltonian causes this symmetry
\cite{Garay,LKO}. For unstable modes the Bogoliubov scalar products
\cite{Fetter,LKO} must vanish,
\begin{eqnarray}
  \int_{-\infty}^{+\infty} (u_n^* u_{n'}-v_n^* v_{n'}) \, dz &=& 0 \, ,
  \label{eq:c1} \\
  \int_{-\infty}^{+\infty} (v_n u_{n'} - u_n v_{n'}) \, dz &=& 0 \, ,
  \label{eq:c2}
\end{eqnarray}
because, in the Bogoliubov-deGennes dynamics
\cite{Dalfovo,Fetter,LKO} the scalar products are stationary
\cite{Fetter,LKO}, whereas the modes (\ref{eq:uv}) are growing or
decaying for complex frequencies. Unstable modes in other field
theories are subject to similar requirements \cite{Unstable}. As a
consequence of the analytic property (\ref{eq:analytic}), condition
(\ref{eq:c1}) is satisfied in the case of purely imaginary
frequencies,
\begin{equation}
  \omega = i \gamma \, .
\end{equation}
Modes (\ref{eq:uv}) with positive $\gamma/\alpha$ are localized near
the horizon at $z=0$. We deform the contour of integral (\ref{eq:c2})
to a large semicircle with radius $r$ around the origin on, say, the
lower half plane, and get
\begin{eqnarray}
  \lefteqn{\int_{-\infty}^{+\infty} (u_n v_{n'} - v_n u_{n'}) \, dz} \nonumber \\
  &\sim& A_n A_{n'} \frac{mc}{\hbar} (\gamma-\gamma') \int_{\pi}^{2 \pi}
  (r e^{i \theta})^{-i (\gamma'+\gamma)/\alpha} \, d \theta \, .
\end{eqnarray}
The integral vanishes for the decay constants
(\ref{eq:result}). Because of the frequency symmetry of the 
Bogoliubov-deGennes equations \cite{Garay,LKO} solutions for negative
$\gamma/\alpha$ must exist as well, although we cannot represent them
as the acoustic Bogoliubov modes (\ref{eq:uv}), because they would
grow in space. Consequently, intrinsic instabilities of sonic
horizons, if any, correspond to the discrete spectrum
(\ref{eq:result}). Let us scrutinize the assumptions made.

Close to the horizon the wavelength of sound would shrink beyond all
scales and the density (\ref{eq:sol}) would diverge, if the wave
equation (\ref{eq:wave}) were universally valid. In the
short-wavelength limit we can describe the Bogoliubov modes in the WKB
approximation \cite{LKO,Csordas},
\begin{eqnarray}
  u_n &=& U_n \exp \left( i \int k \, dz - i \omega t \right) \, , \\
  v_n &=& V_n \exp \left( i \int k \, dz - i \omega t \right) \, .
\end{eqnarray}
We obtain the wavenumber $k$ from Bogoliubov's dispersion relation
\cite{Dalfovo} in moving condensates, taking into account the Doppler
effect,
\begin{equation}
  \label{eq:bogo}
  (\omega- u k )^2 = c^2 k^2 \left( 1+ \frac{k^2}{k_c^2} \right) \, ,
  \quad k_c = \frac{mc}{\hbar} \, .
\end{equation}
The group velocity
\begin{equation}
  \label{eq:v}
  v= \frac{\partial \omega}{\partial k} = u + v' \, , \quad v' =
  c^2 \frac{k}{\omega'} \left( 1+ \frac{2 k^2}{k_c^2} \right) \, ,
\end{equation}
indicates that the acoustic Compton wavenumber $k_c$ defines the
trans-acoustic scale beyond which the excitation velocity in the
fluid, $v'$, deviates significantly from the speed of sound. Consider
the turning point $z_0$ where the group velocity (\ref{eq:v}) vanishes. 
For single-frequency modes,
the excitation flux is conserved \cite{LKO,Csordas},
\begin{equation}
  \partial_z ( U_n^2 - V_n^2) v = 0 \, .
\end{equation}
Consequently, the amplitudes $U_n$ and $V_n$
diverge at the turning point. If the
trans-acoustic scale $k_c$ were zero the horizon itself would be the
turning point. Therefore, we use $|z_0|$ to estimate the spatial range
of the trans-acoustic region. For elementary excitations
\cite{Dalfovo,LKO}, 
\begin{equation}
  \epsilon = \frac{\hbar\omega}{m c^2}
\end{equation}
is a small parameter. We expand $z_0$ in a power series in
$\epsilon^{1/3}$ and solve $v=0$ to leading order \cite{LKO},
\begin{equation}
  \label{eq:z0}
  z_0 \sim \frac{c}{\alpha} \frac{3}{2} \sqrt[3]{-1} \left(
  \frac{\epsilon}{2}\right)^{2/3} \, .
\end{equation}
The trans-acoustic region $|z|\lesssim|z_0|$ is small compared with
$|c/\alpha|$. Therefore, neglecting this region in the integral (\ref{eq:c1})
is justified. Consider the solutions of the dispersion relation
(\ref{eq:bogo}) in the acoustic regime and in the trans-acoustic
extreme. We obtain four branches of this fourth-order equation that
we characterize by their asymptotics on one side of the horizon.
As long as $k^2$ is much smaller than $k_c^2$ we get the
acoustic asymptotics
\begin{equation}
  k \sim \frac{\omega}{k \pm c} \, ,
\end{equation}
and in particular,
\begin{equation}
  \label{eq:kz}
  k \sim \frac{\omega}{\alpha z}
\end{equation}
for sound waves propagating against the current. In the other
extreme, when $k^2$ is much larger than $k_c^2$, one finds
\cite{Corley}
\begin{equation}
  \label{eq:trans}
  k \sim \pm 2 k_c \sqrt{u^2/c^2 - 1} + \frac{ \omega u}{ c^2-u^2} \, .
\end{equation}
Our analysis of the instabilities is justified if the Bogoliubov modes
obey the asymptotics (\ref{eq:kz}) in one of the complex half planes.

\begin{figure}
\includegraphics[width=7cm]{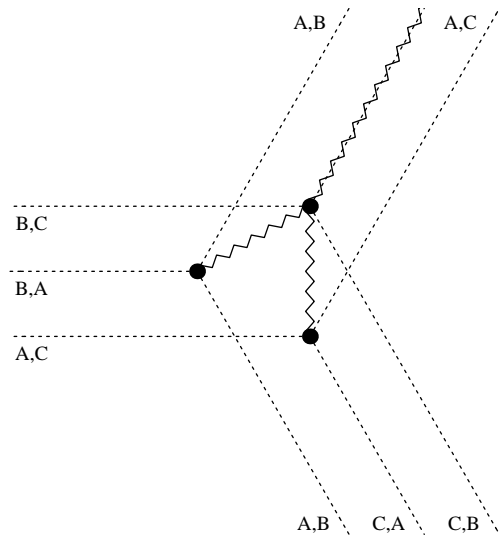}
\caption{\label{fig:stokes1} 
Stokes lines (dotted lines) of unstable elementary excitations 
at a sonic white hole.
The excitations have purely imaginary frequencies.
The Stokes lines originate from the turning points (dots).
The jagged lines indicate the branch cuts of the inverted 
dispersion relation $k=k(\omega)$.
Three branches are connected by the branch cuts and may be converted
into each other at the Stokes lines, the acoustic branch (A) 
with asymptotics (25) and two trans-acoustic branches (B) and (C)
with asymptotics (26).
The pairs of letters indicate which branches of the superposition
are potentially converted into each other. 
The first letter of each pair identifies the exponentially 
dominant branch (determined numerically).
The picture shows that we can construct Bogoliubov modes such 
that the (B) and (C) branches are not present on the lower half plane, 
without causing contradictions.
Therefore, the unstable elementary excitations of the sonic white hole
do indeed obey the asymptotics (25) on the lower half plane.
}
\end{figure}

At a turning point $z_0$ two wavenumber branches of the dispersion
relation (\ref{eq:bogo}) coincide \cite{LKOReview,LKO}. Therefore, a
solution that starts from a particular $k_1$ branch on one side of
$z_0$ may become converted into a superposition of the two modes $k_1$
and $k_2$ with the common turning point. In other words, turning
points may cause scattering. The mode conversion turns out to occur
near specific lines in the complex $z$ plane that are called Stokes
lines in the mathematical literature \cite{Ablowitz}. A Stokes line is
defined as the line where the difference of the WKB phases,
$\int_{z_0}^{z} (k_1-k_2) \, d \zeta$, is purely imaginary. Each of
the three turning points (\ref{eq:z0}) is origin of three Stokes
lines, see the Figure. Where the WKB-phase difference is purely
imaginary, one of the two modes connected by each turning point
is exponentially larger than the other. 
The smaller mode cannot be resolved within the WKB
approximation and may gain a component from the larger mode. The
single-valuedness of the mode function after a complete circle around
the turning points uniquely determines the conversion rules
\cite{Furry}. It follows \cite{Furry} that the exponentially smaller
mode always gains a component from the larger one, if the larger mode
is present. Therefore, to avoid unwanted mode conversion 
the exponentially larger mode should be absent. 
The Figure shows that the elementary
excitations of the sonic white hole can indeed remain on the acoustic
branch (\ref{eq:kz}), which justifies the assumptions made to derive
the result (\ref{eq:result}). Note that the localized acoustic
excitations with positive $\gamma/\alpha$ represent decaying modes for
the white hole when $\alpha$ is negative. The corresponding growing
modes must consist of trans-acoustic excitations with asymptotics
(\ref{eq:trans}). However, as a consequence of the four-fold frequency
symmetry of the Bogoliubov modes \cite{Garay,LKO}, the spectrum
(\ref{eq:result}) is determined by the acoustic modes of the white
hole. In contrast, the unstable excitations of sonic black holes, if any,
cannot possess the asymptotics (\ref{eq:kz}) on one of the complex
half planes. \cite{LKO}. Otherwise, sonic black holes would be always
unstable, and Laval nozzles \cite{Laval,Courant} would be unable to
stabilize fluids that turn from subsonic to supersonic
speed. Apparently, the opposite process, slowing down supersonic
condensates to subsonic speed to form a white hole, is intrinsically
unstable, generating breakdown shocks \cite{Courant}.

In conclusion, 
white holes are best avoided in future experiments to demonstrate
Hawking sound in Bose-Einstein condensates. On the other hand, one
could still use a toroidal geometry \cite{Garay},
as long as the elementary excitations of the torus 
do not match the resonances (\ref{eq:result}). 
Here the condensate should flow through a constriction 
where it exceeds the speed of sound, establishing a black-hole horizon 
followed by a white hole \cite{Garay}.
Because of the periodic boundary condition the spectrum of
excitations is restricted. Our theory seems to explain, at least
qualitatively, why this toroidal arrangement \cite{Garay} exhibits
instabilities at well-defined lines in the parameter space used. 
Instead of employing a torus, one could simply push a condensate
through the optical equivalent of the 
Laval nozzle \cite{Laval,Courant} and let the
supersonic quantum gas expand into space, 
like the solar wind \cite{Parker}.
Otherwise, white holes are
as unstable as wormholes \cite{VisserWorm}.

We thank J. R. Anglin, M. V. Berry, I. A. Brown, J. I. Cirac,
L. J. Garay, T. A. Jacobson, R. Parentani, M. Visser, and G. E. Volovik
for discussions.
Our work was supported by the ESF Programme 
Cosmology in the Laboratory,
the Leverhulme Trust,
the National Science Foundation of Hungary (contract No.\ F032346),
the Marie Curie Programme of the European Commission,
the Royal Society of Edinburgh,
and by the Engineering and Physical Sciences Research Council.


\end{document}